\documentclass[aps,prl,twocolumn,superscriptaddress]{revtex4-1}
\usepackage{amsfonts}
\usepackage{mathtools}
\usepackage{amsmath}
\usepackage{amssymb}
\usepackage{graphicx}
\usepackage{mathrsfs}  
\usepackage{physics}
\usepackage[colorlinks=true,linkcolor=blue,citecolor=red,plainpages=false,pdfpagelabels]
{hyperref}
\usepackage{color}

\allowdisplaybreaks

\providecommand{\U}[1]{\protect\rule{.1in}{.1in}}
\newtheorem{theorem}{Theorem}

\def\Tr{\operatorname{Tr}}

\def\V{\Vert}

\newcommand{\msc}[1]{\mathscr{#1}}
\newcommand{\mc}[1]{\mathcal{#1}}

\begin{document}

\title{Quantum rebound capacity}
\author{Siddhartha Das}\email{sidddas@ulb.ac.be}
\affiliation{Centre for Quantum Information \& Communication (QuIC), \'{E}cole polytechnique de Bruxelles,   Universit\'{e} libre de Bruxelles, Brussels, B-1050, Belgium}
\author{Mark M. Wilde}\email{mwilde@lsu.edu}
\affiliation{Hearne Institute for Theoretical Physics, Department of Physics and Astronomy, Center for Computation and Technology, Louisiana State University, Baton Rouge, Louisiana 70803, USA}

\date{\today}
\begin{abstract}
Inspired by the power of abstraction in information theory, we consider quantum rebound protocols as a way of providing a unifying perspective to deal with several information-processing tasks related to and extending quantum channel discrimination to the Shannon-theoretic regime. Such protocols, defined in the most general quantum-physical way possible, have been considered in the physical context of the DW model of quantum reading [Das and Wilde, arXiv:1703.03706]. In [Das, arXiv:1901.05895], it was discussed how such protocols apply in the different physical context of round-trip communication from one party to another and back. The common point for all quantum rebound tasks is that the decoder himself has access to both the input and output of a randomly selected sequence of channels, and the goal is to determine a message encoded into the channel sequence. As employed in the DW model of quantum reading, the most general quantum-physical strategy that a decoder can employ is an adaptive strategy, in which general quantum operations are executed before and after each call to a channel in the sequence. We determine lower and upper bounds on the quantum rebound capacities in various scenarios of interest, and we also discuss cases in which adaptive schemes provide an advantage over non-adaptive schemes in zero-error quantum rebound protocols. 
\end{abstract}

\maketitle
\textit{Introduction}---One of the great contributions of Shannon was his famous classical channel capacity theorem \cite{Sha48}. A classical channel is described mathematically by a conditional probability matrix $\{p(y|x)\}_{y,x}$, which captures the stochastic nature of a communication medium. Shannon's channel capacity theorem tells us that the ultimate rate at which reliable communication is possible over such a classical channel is equal to the channel's mutual information, which is a function of $\{p(y|x)\}_{y,x}$ that is easy to compute. The power of his mathematical approach is engrained in its abstraction: not only does the theorem apply to a traditional communication setting in which the two communicating parties are spatially separated, but it also applies to a noisy storage scenario in which information can be written to and later read off from a storage device. Thus, from the perspective of information theory, there is no compelling reason to differentiate between these two different physical scenarios, given that the underlying mathematical model can be described in a similar way and the channel capacity theorem is ultimately just a function of the underlying conditional probability matrix $\{p(y|x)\}_{y,x}$.

Many years after Shannon's theory was established and investigated, quantum information theory emerged as a generalization of Shannon's theory, with the main goal being to incorporate the laws of quantum mechanics into Shannon's theory in order to establish the ultimate physical limits of communication (see, e.g., \cite{H06,H13book,W17,Wat16} for reviews of the topic). Interestingly, insights such as teleportation \cite{PhysRevLett.70.1895} and super-dense coding \cite{PhysRevLett.69.2881} led to the realization that there are different kinds of information that can be transmitted over a quantum communication channel, as well as different information-processing tasks \cite{H3LT01,DHW03,DHW05RI,SKBM11,SKBM12,SKBM13}. Again in quantum information theory, the power of the approach taken lies in its abstraction. The various quantum  channel capacity theorems are universally applicable to the processing of arbitrary quantum systems, which include quantum optical systems, superconducting systems, trapped ions, etc.

A particular mathematical model for communication in the quantum setting involves a collection $\{\mathcal{N}^x_{B' \to B}\}_{x\in\msc{X}}$ of quantum channels (in the parlance, each $\mathcal{N}^x_{B' \to B}$ is a completely positive and trace preserving map). The label $x$ in the alphabet $\msc{X}$ indicates a particular channel selected from the collection, and the subscripts $B'$ and $B$ indicate that the same entity (called ``Bob'' here) has access to both the input and output terminals of the channel. It is important to stress that each channel can describe \textit{any physical process} that modifies the quantum system $B'$ input by Bob and returns it back as system $B$, whether it be a noisy storage device or a round-trip communication in which $B'$ goes to another party, who modifies it and returns back to Bob as $B$. Note that the respective input and output systems $B'$ and $B$ need not have the same dimension and could even be labels for quantum systems described by infinite-dimensional Hilbert spaces. The alphabet $\msc{X}$ has cardinality greater than or equal to two and can even be uncountable. As a generalization of quantum channel discrimination \cite{Kit97,AKN97,Aci01,RW05,Sac05,Sac05b,WY06,CDP08a,Hayashi09}, the goal is for Bob to determine a message encoded into a sequence of channels selected from the collection, by employing a quantum physical strategy to do so.

This setting has been studied in various forms and physical instantiations in the literature, starting with the seminal work of \cite{BRV00} from nearly twenty years ago. In the setting of \cite{BRV00}, the collection of channels was restricted to be a collection of unitary channels. The physical scenario considered in \cite{BRV00} was given the name ``communication capacity of quantum computation,'' and the goal is for a computational device to determine a message encoded into one of the unitary channels. Interestingly, it was proposed as an information-theoretic method for determining bounds on the performance of quantum algorithms. Over a decade after the contribution of \cite{BRV00}, the model of quantum reading 
and quantum reading codes were proposed in \cite{Pir11} and \cite{PLG+11,GDN+11}, respectively, inspired by the proposal of quantum illumination from \cite{L08} (see also \cite{Sac05,Sac05b}). In this setting, the collection of channels was allowed to be more general, consisting of completely positive, trace-preserving maps. The physical model considered corresponds to reading out information stored in a classical digital memory, such as a CD or DVD. Quantum physical strategies were allowed, and like what was found in the seminal work on quantum channel discrimination \cite{Sac05,Sac05b}, using entanglement and joint measurements allowed for higher rates of read-out. 

\begin{figure}[ptb]
\begin{center}
\includegraphics[
width=1.0\linewidth
]{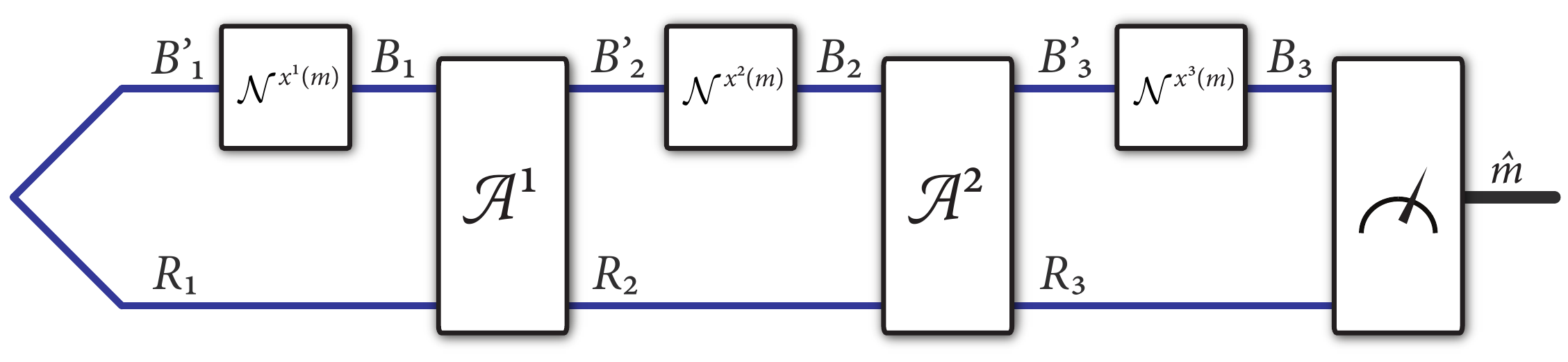}
\end{center}
\caption{The DW model of quantum reading from \cite{DW17}. Beyond the physical setup of quantum reading, this can also be understood in the most abstract sense as a quantum rebound protocol, in which a call to the channel collection or ``channel box'' $\{\mathcal{N}^x_{B' \to B}\}_{x\in\msc{X}}$ is made three times and the most general quantum strategy for decoding the message $m$ is employed.}%
\label{fig:DW-reading-protocol}%
\end{figure}

The work on quantum reading culminated most recently with a general definition of a quantum reading protocol and the related \textit{quantum reading capacity}, given in \cite{DW17} (hereafter, called ``DW model'' of quantum reading). In the DW model of quantum reading, the reader prepares an arbitrary (possibly entangled) quantum state at the start of the protocol, performs adaptive channels between every call to the unknown channels, and finally performs a joint measurement in order to retrieve the encoded message. The DW model of quantum reading is depicted in Figure~\ref{fig:DW-reading-protocol}. This model captures all former and in fact all possible strategies for quantum reading. Several results regarding lower and upper bounds on the rates of quantum reading protocols and quantum reading capacity were established in \cite{DW17}.

It was also stressed in \cite{D18thesis} how the results of \cite{DW17} apply to physical scenarios other than quantum reading, including one in which there is a round-trip communication from one party to another, i.e., in which the channel $\mathcal{N}^x_{B' \to B}$ is implemented physically with the involvement of another party $C$ as $\mathcal{N}^x_{B' \to B} = \mathcal{L}_{C \to B} \circ \mathcal{M}^{x}_{C' \to C} \circ \mathcal{P}_{B' \to C'}$. Here the channel $\mathcal{P}_{B' \to C'}$ describes the forward link from Bob to Charlie, $\mathcal{M}^{x}_{C' \to C}$ describes a local channel that Charlie applies, and $ \mathcal{L}_{C \to B}$ describes the backward link from Charlie to Bob. This kind of setting had been studied previously in \cite{BF02,DL04,S09arXiv,S09,ZZDWS15,ZZDWS16} (in the context of secure communication), and \cite{D18thesis} connected the DW model of quantum reading to this round-trip communication setting.

Inspired by the spirit of abstraction initiated by Shannon in the context of information theory, in this paper a ``quantum rebound protocol'' refers to \textit{any} physical scenario and \textit{any} protocol that decodes information encoded into a collection $\{\mathcal{N}^x_{B' \to B}\}_x$ of channels (the channels can be finite- or infinite-dimensional). This includes all physical scenarios discussed above, i.e., communication capacity of quantum  computation, quantum reading, and round-trip communication protocols. The name is apt, describing exactly how such protocols operate from the perspective of a person who has access to both the input $B'$ and output $B$ of the channel. Indeed, Bob inputs one share of a state into the input port $B'$, the channel $\mathcal{N}^x_{B' \to B}$ is applied, and then system $B$ is returned to Bob, just as it is with a rebound. 

The present paper is a companion to our existing paper \cite{DW17}, with its purpose being two-fold: 1) to clarify that the results of \cite{DW17} should be interpreted in the abstract, information-theoretic way (i.e., as general quantum rebound protocols and not merely as quantum reading ones applied to that physical context), and 2) to discuss in short-paper form the main contributions of~\cite{DW17}.

\textit{Quantum rebound protocol}---Let $\msc{N} \coloneqq \{\mathcal{N}^x_{B' \to B}\}_{x\in \msc{X}}$ be a collection of quantum channels, such that the Hilbert spaces of quantum systems $B'$ and $B$ are described by separable Hilbert spaces. A quantum rebound protocol involves two parties: one party we call ``Alice,'' who selects which message she would like to encode using the channels, and ``Bob,'' who has access to both the input systems labeled by $B'$ and the output systems labeled by $B$. An $(n,R,\varepsilon)$ quantum rebound protocol proceeds as follows:

Both Alice and Bob agree upon message alphabet $\msc{M}$ of size $M$, as well as an $n$-letter codebook (channel sequence) $\{\mc{N}^{x^n(m)}\}_{m\in\msc{M}}$, where
\begin{equation}
\mc{N}^{x^n(m)}\coloneqq (\mc{N}^{x_1(m)},\mc{N}^{x_{2}(m)},\ldots,\mc{N}^{x_n(m)}),
\end{equation}
and $x_{i}(m)\in\msc{X}$. All quantum channels $\mc{N}^x_{B'\to B}$ take states of quantum system $B'$ as input and output states of quantum system $B$. Alice applies the channel sequence $\mc{N}^{x^n}$ based on the message $m\in\msc{M}$ that she wants to communicate to Bob. 

The most general strategy that Bob can adopt for decoding the message $m$ is to transmit a state $\rho_{R_1B'_1}$ through the first call $\mc{N}^{x_1}$ and then perform an adaptive channel $\mc{A}^{(1)}_{R_1B_1\to R_1B_2'}$ after the call $\mc{N}^{x_1}(\rho_{R_1B_1'})$, where $B_i'\simeq B'$ for all $i\in[n]\coloneqq \{1,2,\ldots,n\}$. He then calls the second channel in the sequence, which acts on the output of the previous step. He repeats these steps  until he finishes calling all of the encoding channels in a codeword sequence and finally performs a decoding measurement $\{\Lambda_{R_n B_n}^{(m)}\}_{m\in\mc{M}}$, where $\{\Lambda_{R_n B_n}^{(m)}\}_{m\in\msc{M}}$ is a POVM, i.e., $\sum_{m}\Lambda_{R_n B_n}^{(m)}=I_{R_n B_n}$ and $\Lambda_{R_n B_n}^{(m)}\geq 0$ for all $m\in\msc{M}$. See Figure~\ref{fig:DW-reading-protocol} for a visual depiction of a quantum rebound protocol when $n=3$. Note that the adaptive channels are independent of the codeword sequence and are decided \textit{a priori} based on the codebook. The protocol is such that the average success probability is at least $1-\varepsilon$, for $\varepsilon\in(0,1)$,
\begin{align}
 1- \varepsilon & \leq 1 - p^{(n)}_{\operatorname{err}} \coloneqq 
 \frac{1}{M}\sum_{m}\Tr\left\{\Lambda_{R_n B_n}^{(m)}\rho^{(m)}_{R_nB_n}\right\},\\
\rho^{(m)}_{R_nB_n} & \coloneqq  \mc{N}^{x_n(m)}_{B'_n\to B_n}\circ\mc{A}^{(n-1)}_{R_{n-1}B_{n-1}\to R_nB'_n}\circ \cdots\nonumber\\ 
&\qquad\circ\mc{A}^{1}_{R_1B_1\to R_2B'_2}\circ\mc{N}^{x_1(m)}_{B'_1\to B_1}(\rho_{R_1B'_1}).
\end{align}
The rate $R$ of a given $(n,R,\varepsilon)$ quantum rebound protocol is equal to the number of bits read per channel use:
$
R\coloneqq \frac{1}{n}\log_2 M $.
As emphasized previously, an $(n,R,\varepsilon)$ quantum rebound protocol is no different in the information-theoretic sense from a DW protocol for quantum reading \cite{DW17}, with the only ``difference'' being that the name quantum reading is tied to a particular physical context.

In the above, we have described a non-asymptotic quantum rebound protocol. To go to the asymptotic setting (in the Shannon-theoretic sense), we demand that there exists a sequence of $(n,R,\varepsilon)$ quantum rebound protocols, indexed by $n$, for which $\varepsilon \to 0$ as $n\to \infty$ at a fixed rate~$R$. A rate $R$ is called \textit{achievable} if $\forall \varepsilon\in (0,1)$, $\delta>0$, and sufficiently large $n$, there exists an $(n,R-\delta,\varepsilon)$ quantum rebound protocol. The quantum rebound capacity $C(\msc{N})$ of the channel collection $\msc{N}$ is defined as the supremum of all achievable rates.

As observed in \cite{DW17}, it is apparent that a non-adaptive strategy is a special case of an adaptive strategy, in which the decoder does not perform any adaptive channels and instead uses $\rho_{RB^{'n}}$ as the transmitter state with each $B'_i$ system passing through the corresponding channel $\mc{N}^{x_i(m)}_{B'_i\to B_i}$ and $R$ being an idler system. The final step in such a non-adaptive strategy is to perform a decoding measurement on the composite system $RB^n$.

\textit{Environment-parametrized and environment-seizable collections}---Towards the goal of understanding and establishing limits on quantum rebound capacities, it is of interest to identify quantum channel collections for which we can place an upper bound on their quantum rebound capacities. For environment-seizable channel collections, as defined below, it follows that adaptive strategies do not increase the quantum rebound capacity, such that they can be achieved using non-adaptive strategies. For the larger class of environment-parametrized collections, defined below as well, one can exploit their structure to obtain upper bounds on their quantum rebound capacities~\cite{DW17}.

A collection $\msc{E}=\{\mc{E}^x_{B'\to B}\}_{x\in\msc{X}}$ of quantum channels is called \textit{environment-parametrized} with associated environment states $\{\theta^x_{E}\}_{x\in\msc{X}}$ \cite{DW17} if there exists a fixed channel $\mc{F}_{B'E\to B}$ such that for all input states $\rho_{B'}$, the channel $\mc{E}^x_{B'\to B}$ can be simulated as \cite{TW16} (cf.~\cite{BDSW96,DP05,JWD+08})
\begin{equation}\label{eq:env-collection}
\mc{E}^x_{B'\to B}(\rho_{B'})=\mc{F}_{B'E \to B}(\rho_{B'}\otimes\theta^x_{E}).
\end{equation}

An environment-parametrized collection $\msc{E}$ with associated environment states $\{\theta^x_{E}\}_{x\in\msc{X}}$ is called \textit{environment-seizable} \cite{BHKW18} if there exists a fixed input state $\sigma_{RB'}$ and a fixed channel $\mc{S}_{RB\to E}$ such that for all $x\in\msc{X}$
\begin{equation}
\label{eq:env-seize-collection}
\mc{S}_{RB \to E}(\mc{E}_{B' \to B}^x(\sigma_{RB'})) = \theta^x_{E}.
\end{equation}
In this way, for such environment-seizable channels, one can seize the background environment state $\theta^x_{E}$ with a pre- and post-processing of the channel $\mc{E}_{B' \to B}^x$.  Thus, it is possible to obtain, in a single swoop, the only object $\theta^x_{E}$ distinguishing one channel from another in the collection.

\begin{figure}[ptb]
\begin{center}
\includegraphics[
width=\linewidth
]{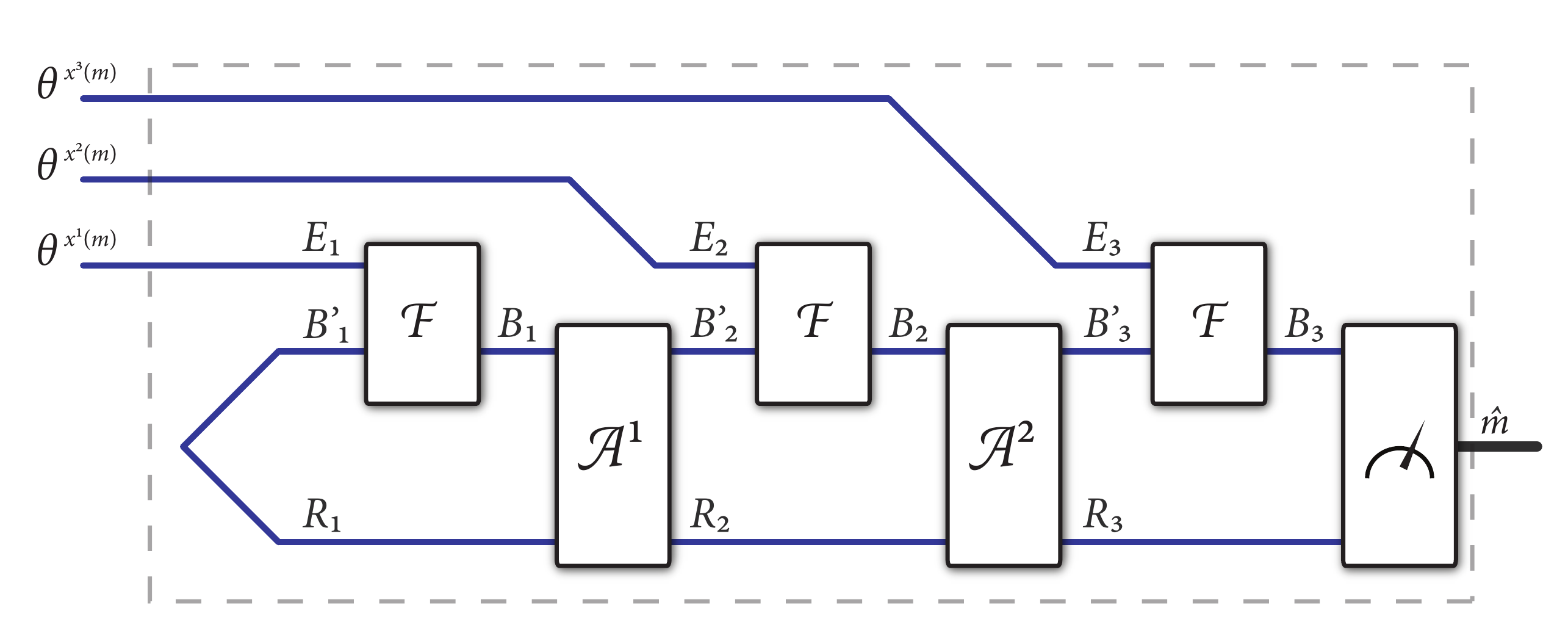}
\end{center}
\caption{The figure depicts how a quantum rebound protocol for an environment-parametrized collection with associated environment states $\{\theta^x_{E}\}_{x\in\msc{X}}$ can be rewritten as a protocol that tries to decode the message $m$ from the environment states $\theta^{x^n(m)}_{E^n}$. All of the operations inside the dashed lines can be understood as a measurement on the states $\theta^{x^n(m)}_{E^n}$.}%
\label{fig:env-param-rebound}%
\end{figure}

\textit{Reduction of rebound protocols for environment-parametrized collections}---In what follows, we show how the structure of general quantum rebound protocols simplify for environment-parametrized channel collections. Let us consider an $(n,R,\varepsilon)$ quantum rebound protocol for an environment-parametrized collection $\msc{E}$ with associated environment states $\{\theta^x_{E}\}_{x\in\msc{X}}$. As shown in \cite{DW17}, the structure of general rebound protocols  simplifies immensely for an environment-parametrized channel collection. This is a consequence of observations made in \cite[Section V]{BDSW96} and \cite{Mul12} in quantum communication theory and in \cite{DM14} in quantum estimation theory. For such an environment-parametrized collection, a quantum rebound protocol can be simulated by one in which every channel use is replaced by a preparation of the environment state $\theta^{x_i(m)}_{E}$ from \eqref{eq:env-collection} and then interacting the channel input with the interaction channel $\mc{F}_{B'E\to B}$. Critically, each interaction channel
$\mc{F}_{B'E\to B}$ is fixed and independent of the message $m\in\mc{M}$. Let 
\begin{equation}
\theta^{x^n(m)}_{E^n}:= \bigotimes_{i=1}^n\theta^{x_i(m)}_{E}
\end{equation}
denote the environment state needed for the simulation of all $n$ of the channel uses in the codeword sequence of the protocol. This leads to the translation of a general quantum rebound protocol to one in which all of the rounds of adaptive channels can be delayed until the very end of the protocol, such that the resulting protocol is a non-adaptive quantum rebound protocol. Figure~\ref{fig:env-param-rebound} displays the reduction.

Thus, any $(n,R,\varepsilon)$ quantum rebound protocol for an environment-parametrized collection $\msc{E}$  can be simulated as a non-adaptive protocol, in the following sense:
\begin{multline}
\!\!\!\! \Tr\left\{\Lambda^{\hat{m}}\left(\mc{E}^{x_n(m)}\circ\mc{A}^{(n-1)}\circ\cdots\circ\mc{A}^{(1)}\circ\mc{E}^{x_1(m)}\right)(\rho_{R_1B'_1})\right\}\\
=\Tr\left\{\Gamma^{\hat{m}}_{E^n}\left(\bigotimes_{i=1}^n\theta^{x_i(m)}_E\right)\right\}, \label{eq:tel-sim-red}
\end{multline}
for some POVM $\{\Gamma^{\hat{m}}_{E^n}\}_{\hat{m}\in \msc{M}}$ that depends only on the choice of the initial state, the adaptive channels, and the final measurement of the original rebound protocol. This adaptive to non-adaptive reduction of a quantum rebound protocol is the same as the adaptive to non-adaptive reduction established in the DW~model of quantum reading \cite{DW17}.  

\textit{Bounds on quantum rebound capacity}---Using the above observation, we now arrive at upper bounds on the performance of any rebound protocol that uses an environment-parametrized collection. Our proof strategy is to employ a generalized divergence \cite{SW12} to make a comparison between the states involved in the actual rebound protocol and one in which the collection $\hat{\msc{E}}:=\{\hat{\mc{E}}_{B'\to B}\}$ of encoding channels has a fixed, single element with environment state $\hat{\theta}_{E}$
and the same interaction channel $\mc{F}_{B'E\to B}$ as in the original collection. The latter rebound protocol contains no information about the message $m$. Also, observe that the augmented collection $\{\msc{E},\hat{\mc{E}}\}$ is environment-parametrized with associated environment states $\{\{\theta^x_{E}\}_{x\in\msc{X}},\hat{\theta}_{E}\}$.

For an $(n,R,\varepsilon)$ quantum rebound protocol that uses an environment-parametrized collection $\msc{E}$, as defined in \eqref{eq:env-collection}, the following upper bound applies to the rate $R$ \cite[Lemma~3]{DW17}:
\begin{equation}
\log_2 M  = nR \leq 
\sup_{p_{X^n}}\inf_{\hat{\theta}} D_h^{\varepsilon}\!(\theta_{X^n E^n}\Vert \hat{\theta}_{X^n E^n}) ,
\label{eq:hypo-test-div-bnd}
\end{equation}
where  $M=|\msc{M}|$ and $D_h^\varepsilon(\rho\Vert\sigma)$ is a generalized divergence called $\varepsilon$-hypothesis-testing divergence \cite{BD10,WR12}, defined for quantum states $\rho,\sigma$ and for $\varepsilon\in[0,1]$ as
\begin{align}
D^\varepsilon_h\!(\rho\Vert\sigma) & \coloneqq 
-\log_2\inf_{\Lambda : 0\leq\Lambda\leq I \wedge\Tr\{\Lambda\rho\}\geq 1-\varepsilon}\Tr\{\Lambda\sigma\},\\
\theta_{X^n E^n}& \coloneqq \sum_{x^n\in\mc{X}^n}p_{X^n}(x^n)\op{x^n}_{X^n}\otimes\theta^{x^n}_{E^n},\\
\hat{\theta}_{X^n E^n}&\coloneqq \sum_{x^n\in\mc{X}^n}p_{X^n}(x^n)\op{x^n}_{X^n}\otimes\hat{\theta}_{E}^{\otimes n},
\end{align}
where $\theta_{X^n E^n}$ and $\hat{\theta}_{X^n E^n}$ are classical--quantum states.

Another generalized divergence of interest is the quantum relative entropy $D(\rho\Vert \sigma)$,  defined for quantum states $\rho,\sigma$ as $
D(\rho\V \sigma):= 
\Tr\{\rho[\log_2\rho-\log_2\sigma]\}$ \cite{Ume62}, which leads to the quantum mutual information $I(A;B)_{\rho}$ for a quantum state $\rho_{AB}$, defined as
$
I(A;B)_{\rho}\coloneqq D(\rho_{AB}\Vert\rho_A\otimes\rho_B),
$
where $\rho_A\coloneqq \Tr_{B}(\rho_{AB})$.

A direct consequence of \eqref{eq:hypo-test-div-bnd} and \cite[Theorem 4]{TT13} is the following theorem (see \cite[Theorem~1]{DW17} for a detailed proof):
\begin{theorem}
\label{thm:env-cell-qrc}
The quantum rebound capacity $C(\msc{E})$ of an environment-parametrized collection $\msc{E}$ with associated environment states $\{\theta^x_{E}\}_{x\in\msc{X}}$, as defined in \eqref{eq:env-collection}, is bounded from above as
\begin{equation}
C(\msc{E})\leq\sup_{p_X}I(X;E)_\theta,
\end{equation} 
where $\theta_{XE}\coloneqq \sum_{x\in\msc{X}}p_{X}(x)\op{x}_X\otimes\theta^x_{E}$ is a classical--quantum state. 
\end{theorem}

\textit{Environment-seizable collections}---When the channel collection is environment seizable, as defined in \eqref{eq:env-seize-collection}, the upper bound in Theorem~\ref{thm:env-cell-qrc} is achievable, so that we have $C(\msc{E})=\sup_{p_X}I(X;E)_\theta$. This equality follows by observing that a strategy for achieving the rate $\sup_{p_X}I(X;E)_\theta$ is to seize every environment state $\theta^x_E$ for each call to the channel, via the pre- and post-processing from the definition in \eqref{eq:env-seize-collection}, and then to employ the achievability part of \cite[Theorem 4]{TT13} (in the asymptotic case, we can invoke the well known result from \cite{Hol98,PhysRevA.56.131}).

A particular example of an environment-seizable collection occurs when the channel collection $\msc{N} = \{ \mathcal{N}^x_{B'\to B}\}_x$ is \textit{jointly covariant}~\cite{DW17}, which results in the environment states being in fact the channel's Choi states $\mc{N}_{B' \to B}^x(\Phi_{RB'})$, where $\Phi_{RB'}$ denotes a maximally entangled state, and the fixed interaction channel is a local operations and classical communication (LOCC) channel $\mc{L}_{B'RB \to B}$,  taken with respect to the bipartition $RB':B$ of the input systems.

Before defining such channel collections, we recall the notion of a covariant channel \cite{Hol02,Hol06,H13book}. Consider a finite group $G$ of size $|G|$. For every $g\in G$, let $g\to U_{B'}(g)$ and $g\to V_B(g)$ be projective unitary representations acting on the input space of $B'$ and the output space of $B$ of a quantum channel $\mc{G}_{B'\to B}$, respectively. The channel $\mc{G}_{B'\to B}$ is covariant with respect to these representations if the following relation is satisfied for all input states $\rho_{B'}$ and for all $g\in G$:
\begin{align}
\label{eq:cov-condition}
(\mc{G}_{B'\to B}\circ \mc{U}^g_{B'})(\rho_{B'}) & = (\mc{V}^g_B\circ\mc{G}_{B'\to B})(\rho_{B'}),\\
\mc{U}_{B'}^g(\cdot) & \coloneqq U_{B'}(g)(\cdot)U^\dag_{B'}(g), \\
\mc{V}_B^g(\cdot) & \coloneqq V_{B}(g)(\cdot)V^\dag_{B}(g).
\end{align}
In this paper, a quantum channel $\mc{G}_{B'\to B}$ is \textit{covariant} if it is covariant with respect to a group $G$ which has a representation $U(g)$, for all $g\in G$, that is a unitary one-design on the channel input system $B'$ ; i.e., the map  $\frac{1}{|G|}\sum_{g\in G}U(g)(\cdot)U^\dagger(g)$ always outputs the maximally mixed state for all input states. 

Finally, a channel collection $\msc{G} = \{\mathcal{G}^x_{B' \to B}\}_x$ is \textit{jointly covariant} if each channel $\mathcal{G}^x_{B' \to B}$ in the collection $\msc{G}$ is covariant with respect to the group $G$. A particular class of jointly covariant channel collections is given by
\begin{equation}
\msc{G}=\left\{\mc{G}_{{B'}\to B}\circ\mc{U}^g_{B'}\right\}_{g\in G}, \label{eq:special-JC}
\end{equation}
 where $\mc{G}_{{B'}\to B}$ is a covariant channel, as defined above. For such jointly covariant channel collections, we have that \cite[Theorem~3]{DW17}

\begin{theorem}\label{thm:covariant-to-EA-cap}
Let $\msc{G}$ be a jointly covariant channel collection as defined in \eqref{eq:special-JC}. Then the quantum rebound capacity $C(\msc{G})$ of the channel collection $\msc{G}$ is equal to the entanglement-assisted classical capacity \cite{BSST99,BSST02} of the underlying quantum channel $\mc{G}_{B'\to B}$:
\begin{equation}
C(\mc{G}) = I(R;B)_{\mc{G}(\Phi)},
\end{equation}
where $\mc{G}(\Phi)\coloneqq \mc{G}_{{B'}\to B}(\Phi_{R{B'}})$ and $\Phi_{R{B'}}$ is a maximally entangled state. 
\end{theorem}

\textit{Zero-error rebound protocols}---Briefly, we mention here that an $(n,R,\varepsilon=0)$ quantum rebound protocol is called a zero-error quantum rebound protocol. By building on results in zero-error channel discrimination \cite{DFY09,HHLW10}, it was shown in \cite{DW17} that there exist channel collections for which a general rebound protocol achieves a higher rate of communication than a non-adaptive one, when it is required for the communication to be zero-error.

\textit{Dense coding capacity}---In a recent posting \cite{L19}, the dense coding capacity was defined in such a way as to generalize earlier work on this topic \cite{H3LT01,DHW03,DHW05RI,SKBM11,SKBM12,SKBM13}. The communication problem formulated there is a particular instance of the DW model of quantum reading \cite{DW17} and thus is immediately seen to be a particular kind of quantum rebound protocol as discussed here. Also, the physical context of round-trip communication for the setting of \cite{L19} was already discussed in \cite{D18thesis} and connected therein to the DW model of quantum reading from \cite{DW17}. The technical contributions of \cite{L19} are Eqs.~(11), (12), and (14) in \cite{L19}, which were established in Theorem~1, Remark~3, and Theorem~3 of \cite{DW17}, respectively. 
 
\textit{Conclusion}---In this paper, we have considered quantum rebound protocols as a way to capture any physical scenario and any information-processing protocol that decodes information encoded into a collection $\{\mathcal{N}^x_{B' \to B}\}_x$ of quantum channels. As done in \cite{DW17} for quantum reading, we have provided a general and natural definition for quantum rebound capacity, by considering that the input and output systems of each channel in the collection are accessible to the same party and arbitrary pre- and post-processing of each channel use is allowed. We have established an upper bound on the quantum rebound capacity for an environment-parametrized channel collection, which is achievable when the channel collection is environment-seizable. We also determined the quantum rebound capacities for jointly covariant channel collections. 

A natural question following from the developments in \cite{DW17} is whether there exists a channel collection for which the quantum rebound capacity is strictly larger than what one could achieve by using a non-adaptive strategy. As discussed above, we have provided a positive answer to this question in the setting of zero error. However, the question remains open for the case of Shannon-theoretic capacity (i.e., with arbitrarily small error). We suspect that this question will have a positive answer, and we strongly suspect it will be the case in the setting of non-asymptotic capacity, our latter suspicion being due to the fact that feedback is known to help in non-asymptotic settings for communication
(see, e.g., \cite{PPV11feedback}). We leave the investigation of this question for future work.

Finally, private rebound protocols consist of communicating information privately via a collection $\{\mathcal{N}^x_{B'\to BE}\}_x$ of quantum wiretap channels. Again, these describe arbitrary physical scenarios in which the decoder has access to both the input and output systems $B'$ and $B$, respectively, while an eavesdropper or wiretapper has access to the system $E$, and the goal is to have reliable communication to the decoder Bob that is private from the eavesdropper. Such protocols were investigated extensively in \cite{BDW18,DBW17}, where lower and upper bounds on communication rates were obtained. The protocols from \cite{BDW18,DBW17} capture not only private reading (defined in \cite{BDW18,DBW17}), but also private round-trip communication protocols such as those discussed in \cite{BF02,DL04,S09arXiv,S09} and floodlight quantum key distribution \cite{ZZDWS15,ZZDWS16}, as discussed in \cite{D18thesis}.

\begin{acknowledgments}
\textit{Acknowledgements}---The research for our companion paper \cite{DW17} was completed when SD was a PhD candidate at Louisiana State University. SD acknowledges support from the F.R.S.-FNRS Foundation under Project No.$~T.0224.18$. MMW acknowledges support from the US Office of Naval Research and the National Science Foundation under Grant No.~1714215. 
\end{acknowledgments}

\bibliographystyle{unsrt}
\bibliography{qr}

\end{document}